\documentstyle[prl,aps,epsf,twocolumn]{revtex}

\input epsf

\newcommand{\cb}{\mbox{$\Xi^{0} \rightarrow
 \Sigma^{+}\, e^{-}\, \overline{\nu}_{e}\,$}}
\newcommand{\clpi}{\mbox{$\Xi^{0} \rightarrow \Lambda\, \pi^{0}\,$}}

\newcommand{\lppi} {\mbox{$ \Lambda  \rightarrow p\, \pi^{-}\,$}}     
\newcommand{\lb}{\mbox{$\Lambda \rightarrow p e^{-}\overline{\nu}_e\,$}}

\newcommand{\csp}{\mbox{$\Xi^{0} \rightarrow \Sigma^{+} \, \pi^{-}\,$}}

\newcommand{\sppi}{\mbox{$\Sigma^{+}  \rightarrow p\, \pi^{0}\,$}}
\newcommand{\spls}{\mbox{$\Sigma^{+}\,$}}

\newcommand{\suf}{\mbox{$ SU(3)_{f}\,$}}
\newcommand{\piz}{\mbox{$\pi^{0}\,$}}
\newcommand{\pim}{\mbox{$\pi^{-}\,$}}
\newcommand{\enu}{\mbox{$e^{-}-\overline{\nu}\,$}}

\newcommand{\nb}{\mbox{$n \rightarrow  p\, e^{-}\, \overline{\nu}_{e}\,$}}
\newcommand{\asp}{\mbox{$ \alpha_{\Sigma^{+}}\,$}}
\newcommand{\gf}{\mbox {$ g_1 / f_1\,$}}
\newcommand{\gfans}{\mbox {$1.32 \pm^{0.21}_{0.17\,{\rm stat}} \pm 0.05_{{\rm syst}}\,$}}
\newcommand{\gfansst}{\mbox {$1.32 \pm^{0.21}_{0.17\,{\rm stat}}\,$}}
\newcommand{\scf}{\mbox {$ g_2 / f_1\,$}}
\newcommand{\scfans}{\mbox {$-1.7 \pm^{2.1}_{2.0\,{\rm stat}} \pm 0.5_{{\rm syst}}\,$}}
\newcommand{\wf}{\mbox {$ f_2 / f_1\,$}}
\newcommand{\wfans}{\mbox {$2.0 \pm 1.2_{{\rm stat}} \pm 0.5_{{\rm syst}}\,$}}
\newcommand{\xpe}{\mbox {$ x_{pe} \,$}}

\newcommand{\xpntr}{\mbox {$ x_{p\nu\perp}\,$}}
\newcommand{\xentr}{\mbox {$ x_{e\nu\perp}\,$}}
\newcommand{\cas}{\mbox{$\Xi^{0}\,$}}

\newcommand{\keth}{\mbox{$K_{L} \rightarrow \pi^{+} e^{-} \overline{\nu}\,$}}
\newcommand{\kethr}{\mbox{$K_{L} \rightarrow \pi^{+} e^{-} \overline{\nu} \gamma\,$}}
\newcommand{\ptsp}{\mbox {$ {\vec p}_{\perp}\,$}}

\newcommand{\lik}{\mbox {$ {\cal L}\,$}}

\newcommand{\qsq}{\mbox {$q^{2}\,$}}

\newcommand{\kef}{\mbox{$K_{L} \rightarrow \pi^{0} \pi^{+} e^{-}\overline{\nu}\,$}}
\newcommand{\kthpi}{\mbox{$K_{L} \rightarrow \pi^{+} \pi^{-} \pi^{0}\,$}}
\newcommand{\asyp}{\mbox{$\alpha_{\Xi} \alpha_{\Lambda}\,$}}

\begin{document}

\title{ First Measurement of Form Factors of the Decay
$\bbox{\Xi^0 \rightarrow \Sigma^+ \: e^- \: \overline{\nu}_e}$
}


\author{
\parindent=0.0in
A.~Alavi-Harati,$^{12}$
T.~Alexopoulos,$^{12}$
M.~Arenton,$^{11}$
K.~Arisaka,$^2$
S.~Averitte,$^{10}$
R.F.~Barbosa,$^{7,\ast}$
A.R.~Barker,$^5$
M.~Barrio,$^4$
L.~Bellantoni,$^7$
A.~Bellavance,$^9$
J.~Belz,$^{10}$
R.~Ben-David,$^{7}$
D.R.~Bergman,$^{10}$
E.~Blucher,$^4$ 
G.J.~Bock,$^7$
C.~Bown,$^4$ 
S.~Bright,$^4$
E.~Cheu,$^1$
S.~Childress,$^7$
R.~Coleman,$^7$
M.D.~Corcoran,$^9$
G.~Corti,$^{11}$ 
B.~Cox,$^{11}$
M.B.~Crisler,$^7$
A.R.~Erwin,$^{12}$
R.~Ford,$^7$
A.~Glazov,$^4$
A.~Golossanov,$^{11}$
G.~Graham,$^{4}$ 
J.~Graham,$^4$
K.~Hagan,$^{11}$
E.~Halkiadakis,$^{10}$
J.~Hamm,$^1$
K.~Hanagaki,$^{8}$
S.~Hidaka,$^8$
Y.B.~Hsiung,$^7$
V.~Jejer,$^{11}$
D.A.~Jensen,$^7$
R.~Kessler,$^4$
H.G.E.~Kobrak,$^{3}$
J.~LaDue,$^5$
A.~Lath,$^{10}$
A.~Ledovskoy,$^{11}$
P.L.~McBride,$^7$
P.~Mikelsons,$^5$
E.~Monnier,$^{4,\dag}$
T.~Nakaya,$^{7}$
K.S.~Nelson,$^{11}$
H.~Nguyen,$^7$
V.~O'Dell,$^7$ 
M.~Pang,$^7$ 
R.~Pordes,$^7$
V.~Prasad,$^4$ 
X.R.~Qi,$^7$
B.~Quinn,$^{4}$
E.J.~Ramberg,$^7$ 
R.E.~Ray,$^7$
A.~Roodman,$^{4}$
M.~Sadamoto,$^8$ 
S.~Schnetzer,$^{10}$
K.~Senyo,$^{8}$ 
P.~Shanahan,$^7$
P.S.~Shawhan,$^{4}$
J.~Shields,$^{11}$
W.~Slater,$^2$
N.~Solomey,$^4$
S.V.~Somalwar,$^{10}$ 
R.L.~Stone,$^{10}$ 
E.C.~Swallow,$^{4,6}$
S.A.~Taegar,$^1$
R.J.~Tesarek,$^{10}$ 
G.B.~Thomson,$^{10}$
P.A.~Toale,$^5$
A.~Tripathi,$^2$
R.~Tschirhart,$^7$
S.E.~Turner,$^2$ 
Y.W.~Wah,$^4$
J.~Wang,$^1$
H.B.~White,$^7$ 
J.~Whitmore,$^7$
B.~Winstein,$^4$ 
R.~Winston,$^{4,\ddag}$ 
T.~Yamanaka,$^8$
E.D.~Zimmerman$^{4}$\\
\vspace*{.1 in} 
(KTeV Collaboration)\\
\vspace*{.1 in}
\footnotesize
{\it
$^1$ University of Arizona, Tucson, Arizona 85721 \\
$^2$ University of California at Los Angeles, Los Angeles, California 90095 \\
$^{3}$ University of California at San Diego, La Jolla, California 92093 \\
$^4$ The Enrico Fermi Institute, The University of Chicago, 
Chicago, Illinois 60637 \\
$^5$ University of Colorado, Boulder, Colorado 80309 \\
$^6$ Elmhurst College, Elmhurst, Illinois 60126 \\
$^7$ Fermi National Accelerator Laboratory, Batavia, Illinois 60510 \\
$^8$ Osaka University, Toyonaka, Osaka 560-0043 Japan \\
$^9$ Rice University, Houston, Texas 77005 \\
$^{10}$ Rutgers University, Piscataway, New Jersey 08854 \\
$^{11}$ The Department of Physics and Institute of Nuclear and 
Particle Physics, University of Virginia,\\ 
Charlottesville, Virginia 22901 \\
$^{12}$ University of Wisconsin, Madison, Wisconsin 53706 \\
}
(Submited to Physical Review Letters, \today)\\
\vspace*{0.1in}
\parbox{14cm}{
We present the first measurement of the form factor ratios
\gf (direct axial-vector to vector), \scf (second class current)  and \wf (weak magnetism)
for the decay \cb using the KTeV (E799) beam line and detector at Fermilab.  
From the \spls polarization
measured with the decay \sppi and the \enu correlation,
we measure \gf to be \gfans, assuming the \suf (flavor) values for \scf 
and \wf .
Our results are all consistent with exact \suf symmetry.
{\flushleft PACS numbers: 13.30.Ce, 14.20.Jn \hspace*{\fill}}
}
\normalsize
}

\maketitle
The study of hyperon beta decay plays a fundamental role in discerning
the structure of hadrons.
The decay \cb is identical to the well measured decay
\nb except that the valence $d$ quarks are replaced by $s$ quarks
in the initial and final state baryons.
In the limit of exact \suf (flavor) symmetry the only differences
between these processes arise from the
different baryon masses and Cabibbo-Kobayashi-Maskawa (CKM) matrix elements.
Modifications to the strong interaction dynamics
due to the difference between the 
$d$ and $s$ quark masses can modify
the form factors from their \suf values.
Different models for \suf symmetry breaking,
using experimental data from other
hyperon beta decays, predict
different values for the
form factors of the decay \cb \cite{rat,fm}.

The general transition amplitude for the semileptonic decay of 
a spin 1/2 baryon (\mbox{$B \rightarrow b\, e^{-}\, \overline{\nu_{e}}\,$}) is:
\begin{eqnarray}
{\cal M } & =  & G_{F}V_{CKM}\frac{\sqrt{2}}{2} \overline{u}_{b}  
( O_{\alpha}^{V} + O_{\alpha}^{A} )
u_{B} \nonumber \\ 
& \times & \overline{u_{e}} \gamma^{\alpha} (1+ \gamma_{5} )
v_{\nu} +  H.c.  ,
\end{eqnarray}
where
\begin{eqnarray}
 O_{\alpha}^{V} & = & f_{1} \gamma_{\alpha} 
+ \frac{f_{2}}{M_{B}}\sigma_{\alpha \beta} 
q^{\beta} +  \frac{f_{3}}{M_{B}}q_{\alpha}, \nonumber \\
 O_{\alpha}^{A} &  = & ( g_{1} \gamma_{\alpha} + \frac{g_{2}}{M_{B}}
\sigma_{\alpha \beta} 
q^{\beta} +  \frac{g_{3}}{M_{B}}q_{\alpha} ) \gamma_{5}, \nonumber \\
q^{\alpha} & = & ( p_{e} + p_{\nu} )^{\alpha} 
= ( p_{B} - p_{b} )^{\alpha}.\  \  
\end{eqnarray}
Here ${G_{F}} $ is the Fermi coupling constant, $ V_{CKM} $ 
is the appropriate CKM matrix element,
and $ M_{B} $ is the mass of the initial baryon.

When $ f_{3} $ and $ g_{3} $ appear 
in the transition amplitude, they are always 
multiplied by the electron mass divided by $ M_B $.  We therefore 
neglect them.  For \cb the predictions from exact \suf symmetry 
(the Cabibbo Model)~\cite{cab} are:
\mbox{$f_{1}=1.0\,$}, \mbox{$g_1=1.27\,$} (from \nb),
\mbox{$f_2=2.6\,$}, \mbox{$g_2=0\,$} (no second class current).
Deviations from exact \suf symmetry 
arising from the differences in the quark masses
can modify the
values of the axial-vector form factors \cite{rat,fm} by up to 20\%.

The KTeV (Fermilab E799) experiment reported the
first observation \cite{cb} of the decay \cb.  
The data presented here were collected
during a later four week period of running in 1997
using an improved trigger.

The KTeV neutral beam is produced 
by an 800~GeV/$c$  proton
beam hitting a 30~cm BeO target at
an angle of  4.8~mrad.  Collimators define two square $0.35\,\mu$sr
secondary beams.  Photons in the beams are converted by a $7.6$~cm
lead absorber, and charged particles are swept 
out of the beam by a series of magnets.
The sweeping magnets also serve to precess the polarization of the 
\cas to the vertical direction.  The polarity of
the final sweeping magnet is regularly flipped
so as to have equal amounts of \cas 
polarized in opposite directions, making the
ensemble average of the \cas polarization negligible.
An evacuated decay volume extends from 94~m to 159~m
downstream of the target.  Downstream of the decay volume is
a charged particle spectrometer consisting of an analysis magnet and
four drift chambers, two upstream and
two downstream of the magnet, 
followed by a CsI electromagnetic calorimeter.
The neutral beams pass through two holes in the calorimeter.  
Other components of the KTeV detector used here are
the photon vetoes
and the system of transition
radiation detectors (TRD).
Details of the detector and trigger system can be found 
elsewhere \cite{cb}.

The decay chain observed here is \cb,
with \sppi and $\pi^{0} \rightarrow \gamma \gamma $.
The final state consists of five particles: a high momentum 
proton which travels through one of the calorimeter beam holes,
a neutrino which is unobserved, an electron
and two photons which are required to hit the calorimeter.

The trigger selects events with a high 
momentum positively charged track (proton) traveling through one of the 
beam holes, an opposite charged track (electron) in the CsI calorimeter, 
and two energy clusters (\piz) not associated with charged tracks.

The decay is reconstructed by finding the
longitudinal position of the \piz decay ($z_{\piz}$)
from the energies and positions of the
photon clusters in the calorimeter using the \piz mass as a constraint.
The photon energies are required 
to be at least 3~GeV and their positions to be at least 1.5~cm 
away from the edge of either beam hole.
The momentum of the \piz is determined from
the extrapolated position of the proton at $z_{\piz}$.
Then the proton and \piz momenta are added to
give the momentum of the \spls.  Finally, the \spls trajectory
is traced back to its closest approach to the electron track,
forming the \cas vertex.

To reduce background from kaon decays, the proton momentum is required to be 
both between 120~GeV/$c$ and 400~GeV/$c$ and greater than $3.6$ times the electron 
momentum.  For electron identification (\pim rejection), we accept only those 
events in which the energy of the calorimeter cluster associated with the 
negative track is within 10\% of the track momentum.  Also, we require a 
\pim probability of less than 0.1 for the TRD signal associated 
with the negative track.

%
%
\begin{figure}[hctbp]
\epsfxsize=8.6cm
\vspace*{-0.8cm}
\epsfbox{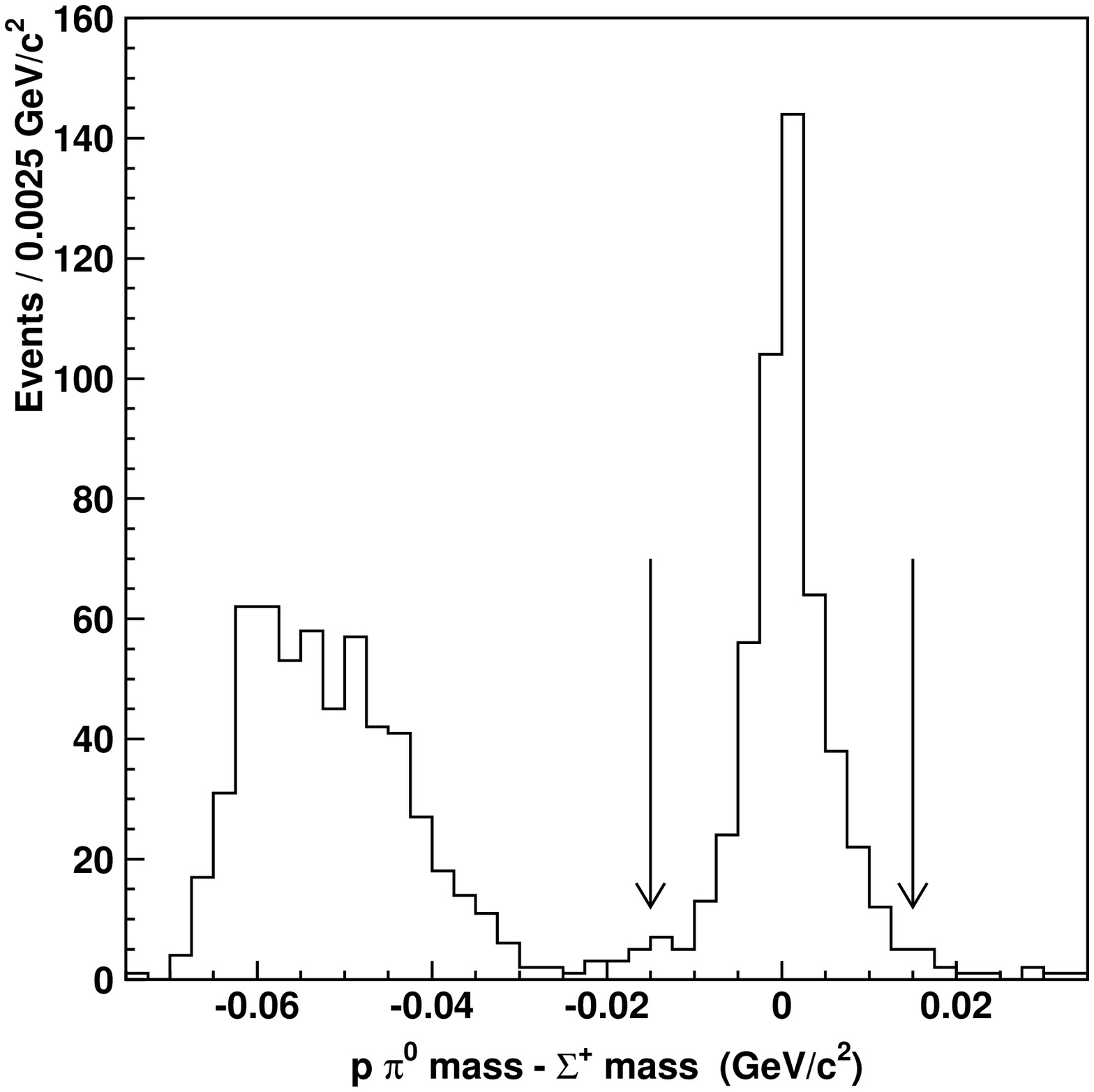}
\caption{ The \sppi mass peak, after all selection 
criteria have been applied.
The background to the left of the peak is due
to \clpi decays (followed by \lppi or \lb).
Since \cb is the only source of \spls
in the beam (\csp is kinematically forbidden), 
signal events are identified by
having a proton-\piz mass within \mbox{15~MeV/$c^2\,$}
of the nominal \spls mass.}
\label{Fig1}
\end{figure}

To remove \kef decays, we require that the 
$\pi^{0} \pi^{+} e^{-}$ invariant mass is greater than \mbox{0.5~GeV/$c^2\,$}, or 
that $z_{\piz}$ is at least 3~m 
downstream of
the \cas vertex.
\kthpi events are suppressed by selecting events with  
a $\pi^{+} \pi^{-} \pi^{0}$  invariant mass greater 
than \mbox{0.57~GeV/$c^2\,$}.  
Photon conversions in the drift chambers upstream
of the analyzing magnet are reduced by rejecting events
with an extra in-time hit in the horizontal views of these chambers.
To reduce background
from \kethr, we reject events with an electron track upstream
segment projected to the CsI calorimeter within
2~cm of a neutral cluster.  For \cb events the \spls vertex is always at
or downstream of \cas vertex within the 1~m measurement error, but for kaon background events
there is no relation between the longitudinal
positions of the observed false
\spls and \cas vertices.
Thus the longitudinal position of
the \spls vertex is required to be no more than
6~m upstream of the \cas vertex, and no
more than 40~m downstream of the \cas vertex.

We calculate the transverse momentum of the 
\cas vertex (\ptsp) by taking the component of the total observed
momentum transverse to the line connecting the target
to the \cas vertex.
Events where the magnitude of \ptsp is larger than the
energy of the neutrino in the \cas frame do not
have a physical solution for the neutrino momentum
and are therefore removed.

Signal events are identified by
having a proton-\piz mass within \mbox{15~MeV/$c^2\,$}
of the nominal \spls mass (Fig.~\ref{Fig1})~\cite{ftnt1}.
After the application of all selection criteria, we
have $494$ events in the signal region.  We estimate
$ 7.4 \pm 3.7 $ background events under the mass peak.
These events are almost entirely due
to \kethr decays with an accidental photon
in the detector (3.4 events), and \keth decays with
two accidental photons in the detector (2.0 events).
We estimate 0.7 background events come from 
\clpi with \lppi and $\pi^{0} \rightarrow e^{+} e^{-} \gamma $,
0.6 events from \kef, and 0.7 events from other sources.

Since the average \cas polarization in our data sample
is negligible, only four kinematic
variables are needed to describe the signal completely.
The process \cb can be described by the
energy of the electron in the \spls frame and the angle between the
electron and neutrino in the \cas frame.
The polarization of the \spls can be described
by the angle between the proton and
the electron,
and the angle between the proton and the
neutrino in the \spls frame.  
The usefulness of the final state polarization
is greatly enhanced by the large asymmetry
of the decay \sppi \, ($\alpha = -0.98$). 

Since the neutrino is unobserved, we cannot unambiguously reconstruct
the directions in the center of mass.  However, assuming the observed
\ptsp is equal and opposite to the transverse momentum of
the neutrino, 
we can obtain unambiguous angular variables transverse to the direction
of the \cas momentum.  Following Dworkin 
\cite{dwo}, we consider the decay sequence
\begin{equation}
\cas \rightarrow Q + \overline{\nu_{e}}\,, \,\,\,Q \rightarrow \spls + e^{-}
\end{equation}
where we have introduced the fictitious particle $Q$.
We then construct angular variables out of these
transverse quantities.  Denoting quantities in the $Q$
rest frame with an asterisk, we have the transverse momenta of the
electron, proton, and neutrino in the $Q$ frame:
\mbox{${\vec{p}}_{e \perp}^{\ *}\,$}, \mbox{${\vec{p}}_{p \perp}^{\ *}\,$} and
\mbox{${\vec{p}}_{\nu \perp}^{\ *}\,$} which is approximately equal 
to \mbox{${\vec{p}}_{\nu \perp}\,$} since the \cas and $Q$ momenta are nearly parallel.

\begin{figure}[hctbp]
\epsfxsize=8.6cm
\vspace*{-0.8cm}
\epsfbox{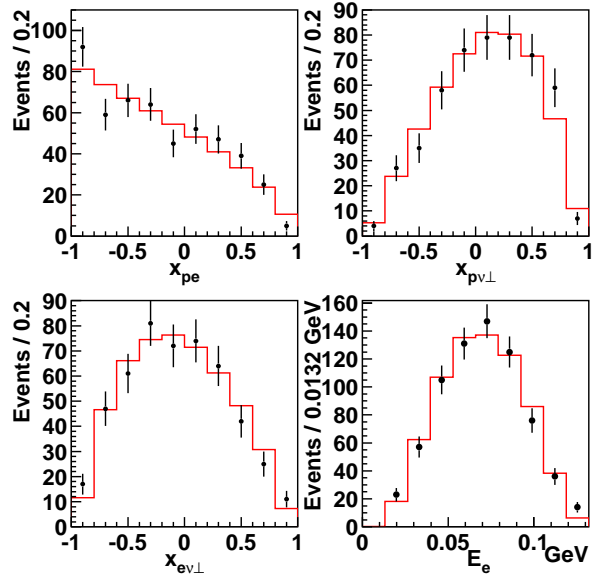}
\caption{ The three variables used to fit \gf and \scf,
and the energy spectrum of the electron in the \spls frame 
(used to determine \wf).  The points
are data and the histogram is our Monte Carlo simulation with
\mbox{$\gf=1.27$} and \mbox{$\scf=0$}. }
\label{Fig2}
\end{figure}

The magnitudes of the momenta in the $Q$ frame are calculated
to obtain the  unambiguous kinematic quantities:
\begin{eqnarray}
\xpntr  =  \frac{{\vec{p}}_{p \perp}^{\ *} \cdot {\vec{p}}_{\nu \perp}}
{\mid  {\vec{p}}_{p}^{\ *} \mid {E_{\nu}}^{*}} 
&\,\,\,\, {\rm and} \,\,\,\,&
\xentr  =  \frac{{\vec{p}}_{e \perp}^{\ *} \cdot {\vec{p}}_{\nu \perp}}
{{E_{e}}^{*} {E_{\nu}}^{*}} 
\end{eqnarray}
which correspond to the polarization of the \spls along the
neutrino direction, and the electron-neutrino correlations,
respectively.  The kinematic quantity corresponding 
to the proton-electron correlation
is \xpe, the cosine of the angle between the proton and the
electron in the \spls frame.  The one dimensional
distributions for \xpe, \xentr and \xpntr are shown in Fig ~\ref{Fig2}.

%
%

\begin{figure}[hhh]
\epsfxsize=8.6cm
\vspace*{-0.8cm}
\epsfbox{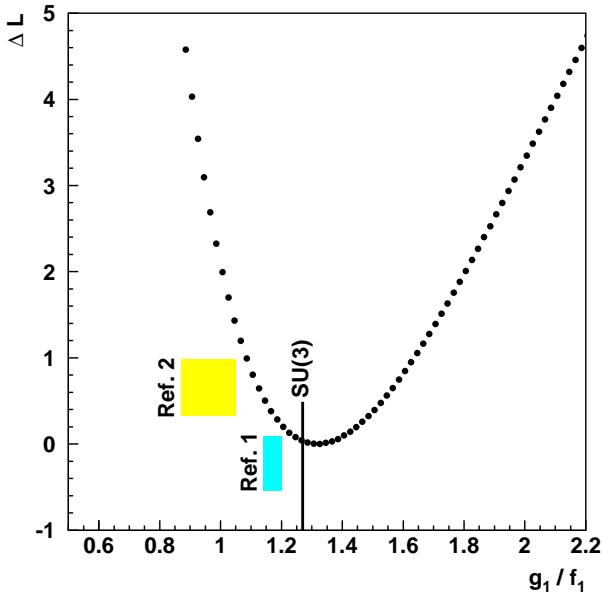}
\caption{ Maximum likelihood fit to \gf.
The shaded bands indicate the range of the theoretical predictions
found in \protect\cite{rat,fm}, the vertical line is the exact \suf value. }
\label{Fig3}
\end{figure}

To determine \gf, rather than calculate the asymmetries 
individually, we perform a maximum likelihood fit
for \gf using \xpe, \xpntr and \xentr.  We create a 
$10 \times 10 \times 10$ histogram for data,
and create corresponding histograms for
different Monte Carlo (MC) values of \gf.
Simulated events are reconstructed 
in the same manner as data events,
and different MC values of \gf
are obtained 
by re-weighting the reconstructed MC events
according to the
differential decay rate \cite{bri,ww} 
using the {\it generated } MC kinematic variables.
We then calculate the log likelihood for each \gf.
The central value is the value of \gf which maximizes the total
log likelihood (\lik ), with the standard errors being determined by
change in \gf which changes \lik\,by 1/2 (Fig.~\ref{Fig3}).
After correcting
for background, 
our final value for \gf is \gfansst.
As a check of our Monte Carlo simulation,
we measure the two body asymmetry product \asyp
with a sample of 70\,000\, \clpi events.
We measure \asyp to be \mbox{$-0.286 \pm 0.008_{{\rm stat}} \pm 0.015_{{\rm syst}}\,$}
which is consistent with its 
value of \mbox{$-0.264 \pm 0.013$}
\cite{pdg}.

The dominant contribution to the systematic error is 
due to the uncertainty in the background.  Other systematic 
errors are estimated by changing quantities in our detector 
commensurate with their
observed deviations from the data (Tab.~\ref{Tab1}).
The systematic error on \gf due to the mass shift is found to be negligible
by comparing MC simulations with \cas masses of \mbox{1.3149~GeV/$c^2\,$}
and \mbox{1.3155~GeV/$c^2\,$}.
\newpage

%
%
\begin{table}[hctbp]
\begin{tabular}{lc}
Source of Uncertainty               & Error on \gf\\
\hline
Background                & 0.039  \\
Beam Shape                & 0.015  \\
MC Statistics             & 0.020 \\ 
DC Alignment                  & 0.020 \\
\cas Lifetime ($\pm 5\%$)   & 0.009 \\ 
CsI Energy Scale ($\pm 0.3\%$)& 0.009 \\ 
Error on \asp ($\pm^{0.017}_{0.015}$)& 0.013 \\ 
\hline
Total  Systematic Error       & 0.054
\end{tabular}
\caption{ Systematic Error for \gf }
\label{Tab1}
\end{table}
\vfill
\vspace*{-1cm}

Radiative corrections have been explicitly determined
not to affect the final state polarization and 
electron-neutrino correlation in hyperon beta decays \cite{gar}.
The standard \qsq 
dependence of $f_1$ and $g_1$ is used \cite{gar},
neglecting the \qsq dependence of $f_1$ and $g_1$ changes the
measured value for \gf by $0.007$.

\begin{figure}[hctbp]
\epsfxsize=8.6cm
\vspace*{-0.8cm}
\epsfbox{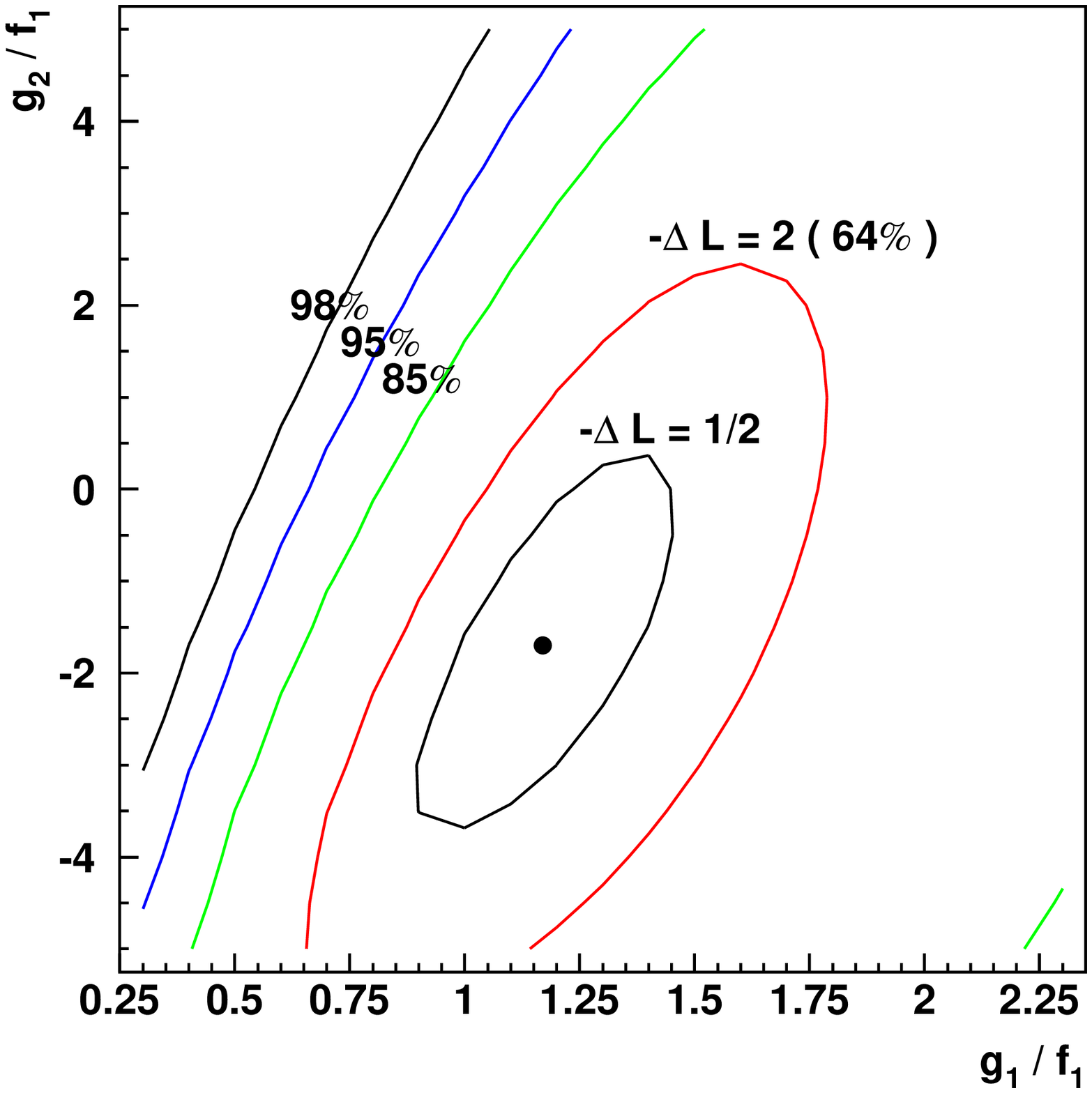}
\caption{ Maximum likelihood fit to \scf and \gf. }
\label{Fig4}
\end{figure}

If we relax the requirement that \mbox{$g_2=0\,$}, and fit the distributions
to \gf and \scf simultaneously, we see no evidence for a
non-zero second class current term (Fig.~\ref{Fig4}),
measuring \mbox{$\gf = 1.17 \pm 0.28_{{\rm stat}} \pm 0.05_{{\rm syst}}\,$}
and \mbox{$\scf =\scfans$}. 

Using our measured \gf, and assuming \mbox{$\scf=0\,$},
we then determine the value for \wf using
the electron energy spectrum in the \spls 
frame (Fig.~\ref{Fig2}). The beta spectrum is the only
kinematic quantity that depends on
\wf to lowest order in \mbox{$(M_{\Xi^{0}}-M_{\Sigma^{+}})/M_{\Xi^{0}\,}$}.
For the \wf measurement, we do not discard events where the
magnitude of \ptsp is greater than
the energy of the neutrino in the \cas frame.
Using a maximum likelihood method, we find the value for \wf is \wfans.

%
%

The systematic errors for \scf and \wf are
determined in a similar manner as those for \gf, however,
a \mbox{0.6~MeV/$c^2\,$} shift in the \cas mass changes \wf by 
0.25, and there is also 
an additional error of 0.3 on \wf due to the statistical error
of \gf.

In conclusion, we have made the first measurement of \gf for the 
decay \cb, and found that \mbox{$\gf = \gfans$} assuming both that no
second class current is present and that the weak
magnetism term has the exact \suf value.
By using
the electron-neutrino correlation and the final state polarization
of the \spls {\it via} its two body decay \sppi,
we are able to determine
both the sign and the magnitude of \gf. Our result agrees well with 
the exact \suf prediction. It therefore favors
\suf breaking schemes that predict small corrections to \gf.
Furthermore,
if we relax the constraint that \mbox{$g_2=0\,$}, and simultaneously fit
for \gf and \scf, we see no evidence for
a second class current term but the uncertainties are large. 
Our analysis of the electron energy
spectrum in the \spls frame
gives a value for \wf that is consistent
with the Conserved Vector Current prediction. 

%
%

We gratefully acknowledge the support and effort of the Fermilab
staff and the technical staffs of the participating institutions for
their vital contributions.  This work was supported in part by the U.S. 
Department of Energy, The National Science Foundation and The Ministry of
Education and Science of Japan.


\begin{references}
\bibitem{ftnt1} We observe that the mean \spls mass is \mbox{$\sim 0.6$~MeV/$c^2\,$}
higher than its nominal value; this shift is also seen in the
\cas mass from \clpi (with \lppi) decays.  Shifting the \spls mass in the
Monte Carlo has a negligible effect on \gf and \scf.
\end{references}
\end{document}